\documentclass[aps,prb,twocolumn,letterpaper,superscriptaddress,showpacs]{revtex4-2}
\usepackage{graphicx}
\usepackage{CJK}
\usepackage{verbatim}
\usepackage{physics}
\usepackage[colorlinks,linkcolor=blue,anchorcolor=blue, citecolor=blue,urlcolor=blue,]{hyperref}
\usepackage{lineno}
\usepackage{bm}
\usepackage{mathptmx}
\makeatletter
\begin{document}
\begin{CJK*}{UTF8}{bsmi}
\title{Relevance of long-range screening in Mott transition examined via a hydrogen lattice}
\author{Zi-Jian Lang (\CJKfamily{gbsn}郎子健)}
\affiliation{School of Physics and Astronomy, Shanghai Jiao Tong University, Shanghai 200240, China}
\author{Sudeshna Sen}
\affiliation{Department of Physics, Indian Institute of Technology (ISM) Dhanbad, India, 826004}
\affiliation{School of Physics and Astronomy, Shanghai Jiao Tong University, Shanghai 200240, China}
\author{Pak Ki Henry Tsang}
\affiliation{Department of Physics and National High Magnetic Field Laboratory, Florida State University, Tallahassee, Florida 32306, USA}
\author{Kristjan Haule}
\affiliation{Center for Materials Theory, Department of Physics \& Astronomy, Rutgers University, Piscataway, NJ, O8854, USA}
\author{Vladimir Dobrosavljevi\'c}
\email{vlad@magnet.fsu.edu.cn}
\affiliation{Department of Physics and National High Magnetic Field Laboratory, Florida State University, Tallahassee, Florida 32306, USA}
\author{Wei Ku (\CJKfamily{bsmi}顧威)}
\email{weiku@sjtu.edu.cn}
\affiliation{School of Physics and Astronomy, Shanghai Jiao Tong University, Shanghai 200240, China}
\affiliation{Ministry of Education Key Laboratory of Artificial Structures and Quantum Control, Shanghai 200240, China}
\affiliation{Shanghai Branch, Hefei National Laboratory, Shanghai 201315, China}
 
\date{\today}

\begin{abstract}
The Mott transition, a metal-insulator transition due to strong electronic interaction, is observed in many materials without an accompanying change of system symmetry.
An important open question in Mott's proposal is the role of long-range screening, whose drastic change across the quantum phase transition may self-consistently make the transition more abrupt, toward a first-order one.
Here we investigate this effect in a model system of hydrogen atoms in a cubic lattice, using charge self-consistent dynamical mean-field theory that incorporates approximately the long-range interaction within the density functional treatment.
We found that the system is well within the charge-transfer regime and that the charge-transfer gap intimately related to the Mott transition closes smoothly instead.
This indicates that the long-range screening does not play an essential role in this prototypical example.
This finding can be understood from the fact that the obtained insulating phase in this model system is driven by strong local interaction, and the transition is associated with the closing of charge-transfer gap.
Contrary to Mott's length scale argument, such energetic competition between kinetic energy and \textit{local} interaction is thus insensitive to long-range screening.
\end{abstract}

\maketitle
\end{CJK*}
\section{Introduction}
Mott transition, as an interaction-driven metal-insulator transition (MIT) without assistance of symmetry change, continues to attract a great attention in the past decades~\cite{Mott1968,Mott1990,Imada}.
It describes the phase transition between a metal and the Mott insulator widely observed in strongly correlated materials, such as transition-metal oxides~\cite{Dagotto,Lee2006,Boer1937,Brito2016,Zhang2017}.
To date, Mott insulators have become one of the most important platforms for the exploration of strongly correlated physics such as unconventional superconductivity~\cite{Dagotto,Damascelli} and magnetism~\cite{Anderson1950}.

The unique characteristic of Mott insulators lies in their unexpected insulating behavior.
Traditional insulators have their Bloch orbitals completely filled by electrons such that a finite energy scale is needed for the electrons to be able to propagate without violating the Pauli principle.
Mott insulators, on the other hand, do not satisfy this condition and therefore would appear metallic in standard band theories, indicating the need to include additional physics beyond the Pauli principle.
To understand such insulating behavior, Mott~\cite{Mott1949,Mott1968} firstly pointed out the importance of electronic interactions, under which the electrons in the system would reside in a bound state and thus unable to propagate without overcoming the effective binding energy.

Following Mott's general picture, modern theoretical studies~\cite{Georges1996,Kotliar2006} of Mott insulators typically make use of half-filled Hubbard model~\cite{Hubbard1963,Hubbard1964_2,Hubbard1964_3,Gutzwiller1963} or its generalized form.
For example, transition-metal oxides~\cite{Boer1937,Brito2016,Zhang2017,Dagotto} and organic Mott insulators~\cite{Powell2011,Pustogow2018} have relatively strong local electronic repulsion $U$ compared to the kinetic energy.
In half-filled systems, such a large $U$ pushes states with doubly occupied orbitals toward high energy and thus allow them to be ``integrated out'' from the low-energy sector, leaving only effectively singly occupied states without charge freedom (thus an insulator.)
From the perspective of Mott's picture, the repulsion $U$ in such half-filled systems provides exactly the necessary electronic interaction to bind the electron in an effective local orbital.

Still, an important physics question remains open, namely the role of the long-range Coulomb interaction on the nature of the quantum Mott transition.
Mott argued~\cite{Mott1949,Mott1968} that between the metallic and insulating phases the significant change in the screening of long-range interaction could lead to a first-order metal-insulator transition at zero temperature.
In contrast, current Hubbard model-based studies (which only incorporate local interaction) found the quantum phase transition to be continuous~\cite{Rozenberg1994_4,Georges1996} instead.
It is therefore an important and timely task to investigate the role of long-range interaction on the Mott transition, in particular to examine the validity of Mott's expectation.

\begin{figure}
	\begin{center}
		\includegraphics[width=\columnwidth]{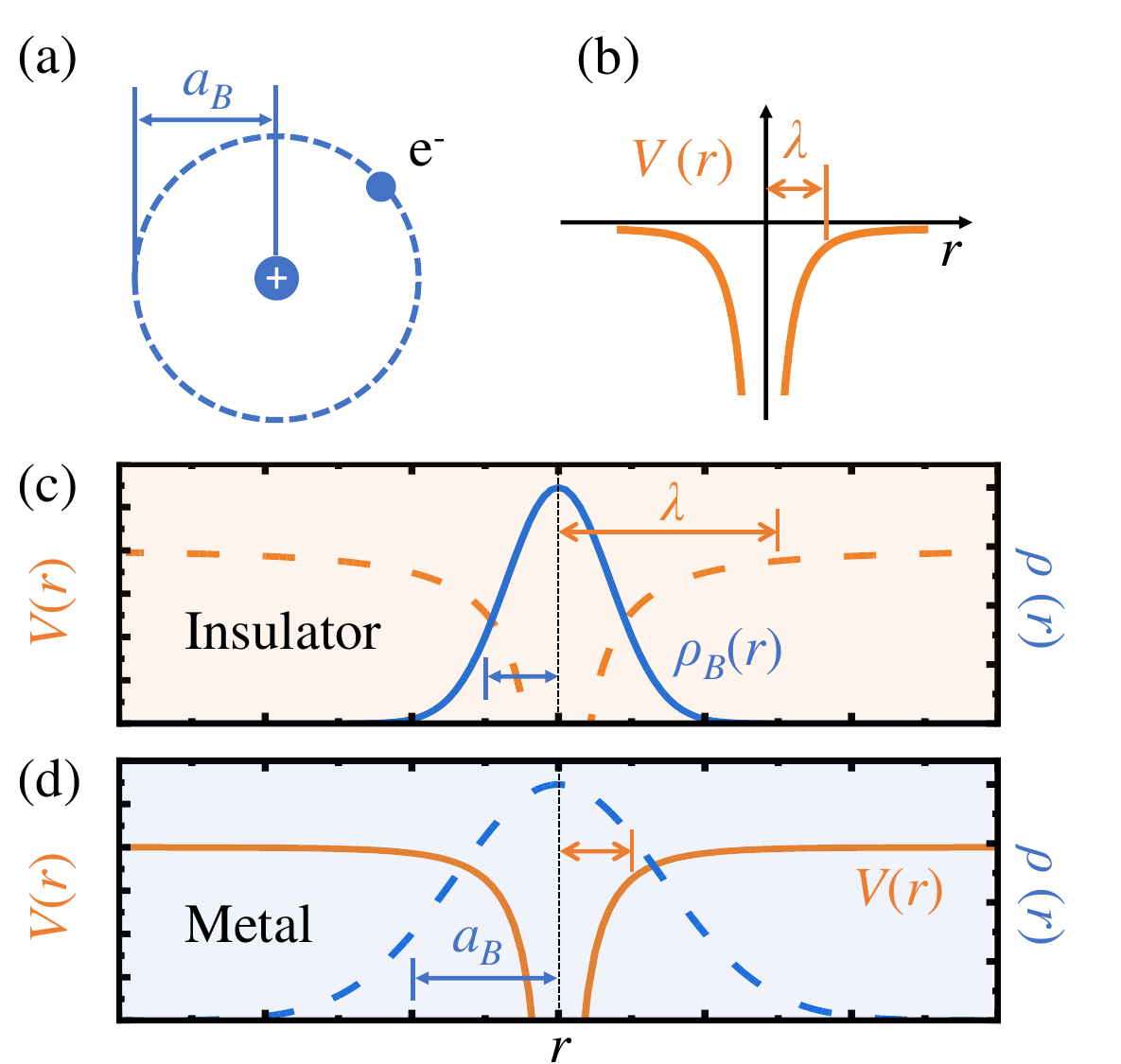}
	\caption{
	Mott's picture of the key scales of metal-insulator transition: (a) radius $a_B$ of the particle-hole bound state, and (b) range $\lambda$ of screened long-range Coulomb attraction $V(r)$ as a function of relative distance $r$.
	(c) When the bound state density $\rho_B$ is confined within the Coulomb attraction, $a_B < \lambda$, the system is insulating.
	(d) Otherwise the system becomes metallic.
	The finite $\lambda$ at the transition dictates a first-order phase transition in this picture.
	}
	\label{MIT_fig1}
	\end{center}
\end{figure}

Here, to address this essential question, we attempt to incorporate the effect of long-range interaction and investigate the Mott transition of a model system  with hydrogen atoms in a cubic lattice.
The effect of interaction screening is incorporated approximately via density self-consistency~\cite{Haule2010,Haule2015} of the density functional theory plus dynamic mean-filed theory (DFT+DMFT)~\cite{Hohenberg1964,Kohn1965,Metzner1989,Muller1989,Brandt1989,Janis1991,Georges1992,Georges1996,Kotliar2006}.
We found that this model system is in the charge-transfer regime, in which the relevant low-energy charge sector is dominated by the 2$s$ and singly occupied 1$s$ orbitals.
Correspondingly, the Mott transition is intimately related to the closing of the charge-transfer gap.
Importantly, the insulating gap is found to reduce \textit{smoothly} to zero and the charge carrier density (reflected by $1s$ orbital occupations) grows continuously.
Both results suggest a continuous quantum phase transition, instead of the first-order transition expected by Mott's consideration of long-range screening.
Our result can be understood from the fact that this common type of Mott transition in materials is mainly controlled by competition between kinetic and \textit{local} interaction energies.
Contrary to Mott's \textit{length} scale argument, such an \textit{energetic} competition is therefore insensitive to screening of long-range interaction.

\begin{figure}
	\setlength{\belowcaptionskip}{-0.4cm}
	\begin{center}
	\includegraphics[width=\columnwidth]{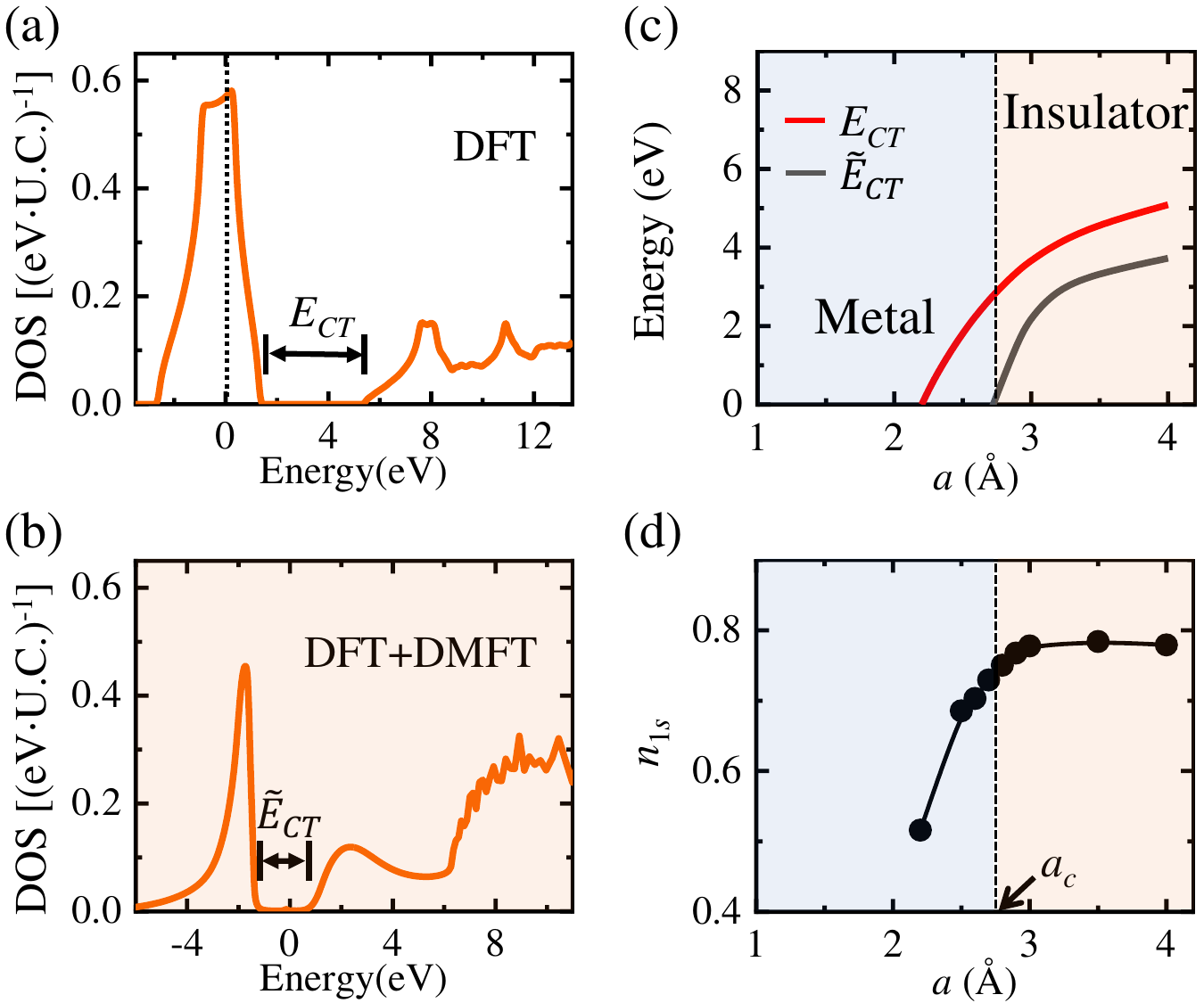}
	\end{center}
	\caption{
	(a) One-body density of states (DOS) in unit of (eV$\cdot$ unit cell)$^{-1}$ for $a=3.0\text{\AA}$ via standard DFT treatment, indicating a metallic system with a charge-transfer gap $E_{CT}$ above the chemical potential set as the reference energy at zero.
	(b) The same via DFT+DMFT at $T=0.01~$eV, giving an insulating system with a correlation-renormalized charge-transfer gap $\tilde{E}_{CT}$ across the chemical potential.
	(c) Smooth reduction of charge-transfer gap $E_{CT}$ and $\tilde{E}_{CT}$ as the lattice constant $a$ decreases toward the quantum critical point at $a_c$, below which $\tilde{E}_{CT} < 0$ and the system turns metallic with a gradual increase of itinerant hole carrier density reflected by the smooth reduction of $1s$ occupation within a fixed atomic sphere (d).
	Both (c) and (d) indicate a continuous quantum phase transition different from Mott's proposal.
	}
	\label{MIT_fig2}
\end{figure}


\section{Mott's Picture}
Let's first review the original consideration of the Mott transition.
Mott proposed that a lattice of one-electron atom, such as hydrogen, must be an insulator at large lattice spacing.
This is because starting from a ground state with a uniform charge of one electron per atom, the low-energy current response must be gapped associated with the formation of particle-hole bound states with an effective Bohr radius $a_B$ [cf. Fig.~\ref{MIT_fig1}(a)] (assuming a poorly screened $e^2/r$ attractive Coulomb interaction over distance $r$.)
On the other hand, at small lattice spacing the system is expected to become a good metal with a half-filled lowest energy band.
This is because the current response now turns gapless in association with the delocalization of particle-hole pairs, as a result of the long-range screening that reduces the range of the attraction $-(e^2/r) \text{exp}(-2\pi r/\lambda)$ to a finite length scale $\lambda$ [cf. Fig.~\ref{MIT_fig1}(b)].
Therefore, in such a \textit{length} scale consideration, there must exist a transition point  for a metal-insulator transition at some intermediate lattice spacing.

As a simple estimation, Mott proposed that the metal-insulator transition would take place at the point when $a_B$ and $\lambda$ are equal.
This seems quite reasonable since at $a_B \ll \lambda$ [cf. Fig.~\ref{MIT_fig1}(c)] the bound state is well preserved such that the system should be a good insulator, while at  $a_B \gg \lambda$ [cf. Fig.~\ref{MIT_fig1}(d)] the interaction is screened too strongly to maintain a bound state such that the system would be a good metal.
Based on this rough criteria, Mott estimated the critical density $n_c^{1/3} a_B \sim 0.25$~\cite{Mott1956} via Thomas-Fermi approximation. 
Though this estimation is very crude, the criteria actually works very well for many materials~\cite{Mott1990}.

The necessity of a finite $\lambda \sim a_B$ in Mott's criterion of metal-insulator transition suggests strongly that the transition be first-order, since a finite $\lambda$ corresponds to a finite itinerant carrier density.
As another way to visualize this, near the critical density, a slight increase of itinerant electrons would lead to an enhanced screening (or reduced $\lambda$), which in turn further increases the carrier density.
One would therefore expect that such a non-linear feedback can lead to a discontinuous transition at zero temperature.
Unfortunately due to the complexity of quantum many-body problems, theoretical studies to date have not been able to incorporate the screening of the length scale of long-range interactions, nor has the corresponding first-order quantum phase transition been obtained~\cite{Georges1996,Kotliar2006}.

\section{Method}
To include this important screening effect of long-range Coulomb interaction beyond the Thomas-Fermi approximation, we use the density self-consistency scheme~\cite{Haule2010,Haule2015} of density functional theory (DFT) plus dynamical mean-field theory (DMFT)~\cite{Karp2020,Pajskr2016,Craco2019,Xie2022,Brito2016,Leonov2015}, with previously established double counting formulation~\cite{Haule2015}.
In the process of self-consistent iteration, the change of electron density and corresponding screening effect will feedback to the next iteration and thus allow the possibility of the first-order transition.
The strong intra-atomic repulsion $U$ ($\sim1$Ry~\cite{Unote,Nakamura2021,Giannozzi2020}) is included via DMFT~\cite{Metzner1989,Muller1989,Brandt1989,Janis1991,Georges1992,Georges1996,Kotliar2006} which is so far the most widely applied approximation to achieve the Mott insulating phase.
This scheme is implemented by DFT+embedded DMFT functional~\cite{Haule2010,Haule2015} (See Appendix~\ref{cal_details} for calculation details).

\section{Results}
We first start with a low-density insulating case at lattice spacing $a=3.0~\text{\AA}$.
Fig.~\ref{MIT_fig2}(a) shows the resulting density of states (DOS) in the standard DFT calculation via local density approximation~\cite{Hohenberg1964,Kohn1965}.
As expected, the system is incorrectly identified as a metal with chemical potential in the middle of the first band, corresponding to the half-filling $1s$ orbital.
Upon inclusion of strong local exchange-correlation via DMFT at $T=0.01~$eV, Fig.~\ref{MIT_fig2}(b) shows the correct insulating state with a DOS hosting a gap around the chemical potential.

It is obvious that this insulating gap is not the standard Mott gap in a Hubbard model, since its size(cf. Appendix~\ref{cal_details}) is only around $2~$eV in this case, much smaller than the local repulsion $U\sim 1~$Ry.
Indeed, in both DFT [panel (a)] and DFT+DMFT [panel (b)] results, the orbital content right above the gap is predominately of 2$s$ and 2$p$ instead of 1$s$.
Therefore, within the parameter range near the Mott transition, this model system is well within the charge-transfer (CT) insulator regime, in which the lowest-energy charge fluctuation is between 1$s$ and 2$s$/2$p$ orbitals of neighboring atoms.


Notice that the charge-transfer gap $E_{CT}$ is strongly reduced from $4~$eV in the DFT result [panel(a)] to only $\sim2~$eV in the DFT+DMFT result [panel(b)].
This can be understood from the additional energy lowering from the emergent antiferromagnetic spin exchange between the itinerant carrier in 2$s$/2$p$ and the local immobile 1$s$ spin, similar to the one that drives the formation of Kondo singlet~\cite{Kondo1964,Ng1988}.

Now approaching the metal-insulating transition via reducing the lattice constant, as shown by the black line in Fig.~\ref{MIT_fig2}(c), the CT gap of the insulating solution reduces and eventually vanishes around $a=a_c\sim 2.8~\text{\AA}$.
Naturally, after that the insulator solution can no longer be found and the system is metallic.
Incidentally, for $a > a_c$ metallic solution is unstable as well.
This indicates that the phase transition is intimately related to the gap closing.
(Note that in an extremely narrow range of $\Delta a \sim 0.003~\text{\AA}$ around $a_c$, both metallic and insulating solutions can be found[cf. Append.~\ref{entropy_1st}].
Considerations on this nearly negligible and somewhat artificial feature will be discussed later.)

Most importantly, our results display a very important deviation from Mott's expectation on the phase transition.
Figure~\ref{MIT_fig2}(c) shows that even with the long-range Coulomb interaction and its screening incorporated via the DFT treatment, as the lattice spacing decreases, $\tilde{E}_{CT}$ only smoothly reduces to zero.
Furthermore, as the lattice spacing decreases further, Fig.~\ref{MIT_fig2}(d) shows that the occupation of the $1s$ orbital (within a fixed atomic sphere~\cite{Blaha2020} of radius 1.8 a.u.) decreases smoothly as well, indicating a gradual development of hole carrier density.
No abrupt enhancement of gap closing or the hole carrier density is found at $a_c$ that corroborates Mott's expectation of a first-order phase transition.
Clearly, our results suggest a continuous metal-insulator transition.
In other words, within the approximation of the state-of-art DFT+DMFT approach (omission of non-local correlation), the proposed drastic effect of screening of long-range interaction does not appear to be essential to the Mott transition in this model system.

\section{Discussion}
This lack of relevance of long-range screening can be understood from the following energetic considerations of this model system.
In this charge transfer insulator, mobile carriers can only occur in the ground state when the fully renormalized bandwidths grow to overcome the orbital energy and eliminate the \textit{indirect} normalized charge-transfer gap minimum, $\tilde{E}_{CT}$, [between momentum ($\pi$,$\pi$,$\pi$) and ($\pi$,0,0)].
However, this system has a rather large orbital energy difference ($\sim10$ eV) between the dressed 2s and 1s orbitals and a large kinetic energy represented by the bandwidths of the dressed 1s band ($W_{1s}\sim 5$ eV) and the dressed 2s band ($W_{2s}\sim 20$ eV).
Both of these key high-energy factors are insensitive to long-range interaction and thus its screening.
Particularly, no dramatic change of long-range screening (of wavevector $q\sim 0$) is expected even right before the \textit{indirect} gap is closed, since the direct gap remains very large, of 10 eV scale.

In real materials, even beyond this point of gap closing, the mobile electrons in the dressed 2s orbital and holes in the dressed 1s orbital may still be bound into charge-neutral excitons by the long-range interactions.
To such an excitonic insulator~\cite{Kohn1967,Kohn1967_2}, Mott's consideration of the insulating phase in Fig.~\ref{MIT_fig1}(c) may apply, in which the length scale of the long-range interaction is longer than the size of the bound excitons.
Nevertheless, given that such excitonic binding should be of the order of 100 meV at best, two orders of magnitude weaker than the above kinetic energies, such an excitonic insulating phase, if present, must only delay the emergence of metallic carriers in a negligibly narrow region.
Consequently, only a small number of carriers would be present when entering the metallic phase.
Furthermore, this excitonic effect would be greatly suppressed by the dramatic difference in the effective masses of the electrons and holes.
Therefore, even if Mott's proposal of a first-order transition may be realized here, the effect is likely too weak for the accuracy of realistic many-body calculations to date.
On the technical side, such excitonic physics that involves non-local particle-hole pairs is surely beyond the capability of the LDA+DMFT framework employed in this study, or even the advocated {\it GW}+EDMFT framework~\cite{Biermann2003,Ayral2013,Hansmann2013,Petocchi2020} as well.

Our conclusion and the analysis above are qualitatively different from those of the previous extended DMFT studies on extended Hubbard model~\cite{Kotliar2000,Biermann2003,Ayral2013,Rubtsov2012,Hansmann2013,Huang2014,van2014,Hafermann2014, Petocchi2020,Rohringer2018,Schuler2019,Terletska2021}, which found a first-order quantum phase transition instead~\cite{Kotliar2000}.
Note that those results originate from the non-linear feedback in mapping the long-range Coulomb interaction onto retarded (dynamical) screening of \textit{local} $U$ of the Kondo impurity model.
Such mechanism is therefore physically distinct from (and unable to address) Mott's proposal of screening the \textit{length scale} of long-range interaction.
In contrast, the key factor for the MIT here is the closing of the \textit{indirect} charge transfer gap, which is insensitive to the long-range screening.
Correspondingly, a smooth continuous transition naturally results in our case.

Finally, it is well known~\cite{Georges1996,Kotliar2006,Vollhardt2012} that the lack of interatomic correlation in DMFT treatments can lead to an unphysically large degree of degeneracy and in turn an artificially large entropy at finite temperature $T=0^+$.
This often causes an entropy-driven first-order transition at finite temperature~\cite{Georges1993,Rozenberg1994}.
Our finite-temperature calculation and the effective analytical model above are not immune to this problem(cf. Appendix.~\ref{entropy_1st}).
Nonetheless, it is interesting to note that this artifact is much weaker in this model charge-transfer systems than that in the Hubbard model.
In fact, in Fig.~\ref{MIT_fig2}(c), the region hosting first-order transition is narrower than the thickness of the black line.
We take it that this distinction reflects the more robust continuous nature of the quantum phase transition in such charge-transfer systems.

Experimentally, our model system of periodic hydrogen lattice could be tested by the next generation of capabilities that engineers hydrogen-like lattice of any desired geometry through STM-implanting donors on surfaces of semiconductors~\cite{Schofield2013}.
These experimental advances allow tuning the system across Mott transition without randomness and are thus ideal to experimentally clarify the role of long-range interaction.

\section{Conclusion}
In summary, we investigate the important open question concerning the role of long-range screening on the quantum Mott transition using a model system of hydrogen lattice.
We employ the charge self-consistent scheme of the dynamical mean-field theory that incorporates approximately the long-range interaction and its screening within the density functional treatment.
The system is found to be well within the charge-transfer regime and the charge-transfer gap closes smoothly across the quantum phase transition.
Together with a gradually increasing hole carrier density in the metallic phase, this indicates a continuous quantum phase transition distinct from Mott's proposal.
This deviation from Mott's long-range screening-based first-order picture can be understood from 1) the obtained insulating phase in this typical case is driven by short-range interaction, and 2) the metal-insulator transition involves the closing of the screening-insensitive charge-transfer \textit{energy}, instead of \textit{length}-scales switching proposed by Mott.
Quantum metal-insulator transitions in such charge-transfer materials are thus insensitive to the long-range interaction.
Our study leaves open a finite possibility of weak excitonic insulating phase in a very narrow region, in which Mott's picture might still apply.
Further resolution of this possibility would require a significant advance in current computational capability.

\begin{acknowledgments}
    We thank Yuting Tan and Xinyi Li for their helpful discussions.
This work is supported by National Natural Science Foundation of China (NSFC) under Grants No. 12274287 \& 12042507, and Innovation Program for Quantum Science and Technology No. 2021ZD0301900. 
S.S. acknowledges support from the Science and Engineering Research Board, India
(SRG/2022/000495), (MTR/2022/000638), and IIT(ISM) Dhanbad [FRS(175)/2022- 2023/PHYSICS].
Work in Florida (V.D. and P.K.H.T.) was supported by the NSF Grant No. DMR-2409911, and the National High Magnetic Field Laboratory through the NSF Cooperative Agreement No. DMR-2128556 and the State of Florida.
K.H. was supported by NSF DMR-2233892.
\end{acknowledgments}

\appendix
\section{Calculation details}
\label{cal_details}
Our self-consistent DFT+DMFT calculation employs an implementation~\cite{Haule2010,Haule2015} of the eDMFT method~\cite{Georges1996,Kotliar2006} based on the WIEN2k package~\cite{Blaha2020}.
The double-counting choice is `exact'~\cite{Haule2015}.
We set a model system with a simple cubic hydrogen lattice with various lattice constants and a fixed on-site interaction $U$=1~Ry on the $1s$ orbital. 
With the decrease of lattice constant, the kinetic hopping between atoms increases. 
A large energy window [-30eV, 30eV] is used in defining the quantum Hilbert space that provides channels for various hopping and charge-transfer processes.
The real-frequency dependence of the density of states is obtained from analytical continuation via maximum entropy method~\cite{analy_cont} from the Matsubara-frequency Green's function.

\begin{figure}[b]
	\setlength{\belowcaptionskip}{-0.4cm}
	\begin{center}
	\includegraphics[width=\columnwidth]{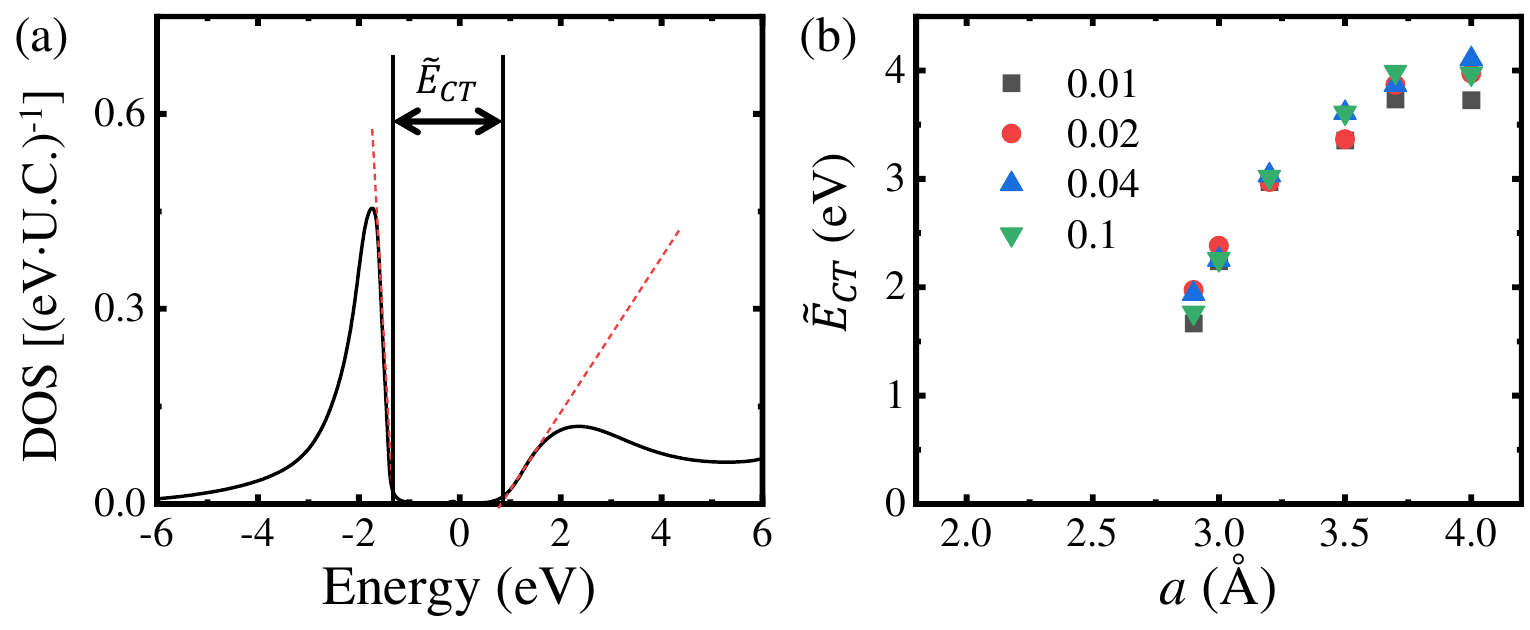}
	\end{center}
	\caption{(a)Determination of charge-transfer gap (insulating gap) $\tilde{E}_{CT}$ in DFT+DMFT calculation. Red dashed line is the linear extrapolation near the edge of the gap.
 (b) Charge-transfer gap $\tilde{E}_{CT}$ as a function of lattice spacing $a$ at temperature 0.1 (green triangle), 0.02 (blue triangle), 0.04 (red circle) and 0.01 (black square)~eV.
	}
	\label{figs1}
\end{figure}

The bare charge-transfer gap $E_{CT}$ is determined from the density of states (DOS) of the DFT calculation where the value of DOS goes to zero.
This however is not applicable in the DFT+DMFT calculations in which the DOS are obtained by the maximum entropy method~\cite{analy_cont} such that DOS can never reduce to absolute zero.
The charge-transfer gap (insulating gap) $\tilde{E}_{CT}$ in DFT+DMFT calculation shown in Figure 1(c) is determined by the linear extrapolation near the gap edge as shown in Fig.~\ref{figs1}(a).
Since the gaps are estimated in the regime far away from the phase transition, they are barely influenced by the temperature which is much smaller compared to these gaps in the energy scale as shown in Fig.~\ref{figs1}(b).
One can find that the gaps remain almost the same upon increasing temperature from 0.01 to 0.1 eV.

\section{Entropy driven first-order phase transition}
\label{entropy_1st}

\begin{figure}
	\begin{center}
	\includegraphics[width=\columnwidth]{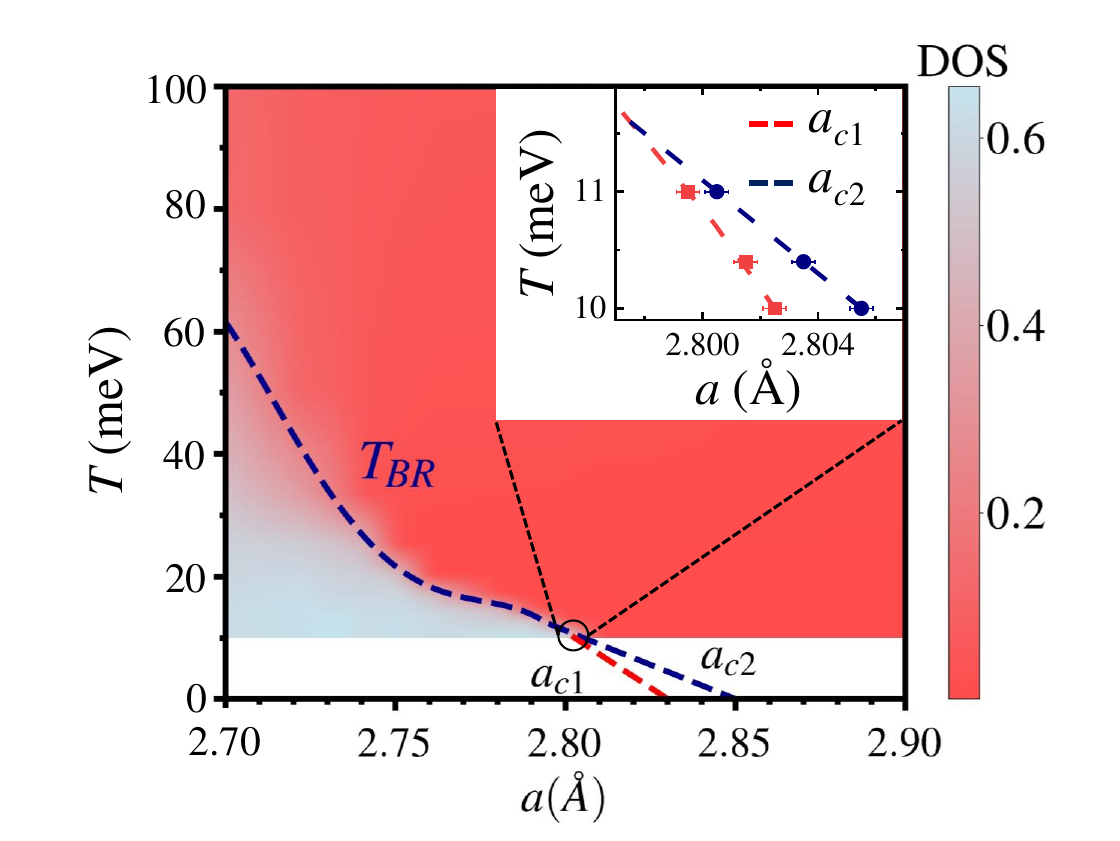}
	\end{center}
	\caption{Phase diagram of real hydrogen lattice as a function of temperature and lattice constant. The color map shows the value of DOS at chemical potential. Two dashed lines are the boundary of the coexistent region. The dark blue line $T_{BR}(a) $ is the Brinkman-Rice line marking the thermal destruction of metallic quasiparticles~\cite{Brinkman1970}.
	The inserted figure is the coexistent region with two boundaries $a_{c1}(T)$ and $a_{c2}(T)$ denoted by red and dark blue dashed lines respectively.
	}
	\label{MIT_fig3}
\end{figure}
It is well known that DMFT calculation at finite temperature suffers from the lack of non-local correlation, which introduces a large entropy contribution of the local spin even at $T=0^+$~\cite{Vollhardt2012}.
For the particular issue of Mott transition, such a large entropy is known to lead to a first-order phase transition at finite temperature calculation~\cite{Georges1996,Kotliar2006}, even though the quantum phase transition is continuous.
For the typical one-band case, there is a coexistent region which suggests a first-order phase transition.
In our model system, if one carefully tunes the lattice space, the coexistent region can be still found in an extremely tiny region.
As the inserted figure shown in Fig.~\ref{MIT_fig3}, we also find a coexistent region below $T\sim10$~meV with an extremely narrow size which demonstrates a first-order phase transition at finite temperature.
This phase transition is however driven by the entropy instead of the screening effect suggested by Mott.
More importantly, this artificially large entropy does not affect the continuous nature of the quantum phase transition.

It's interesting to note that compared with the coexistent region of the typical one, there are two main difference. 
First, the value of critical on-site repulsion $U_c$ is much larger.
Usually, $U_c\sim 2D$ is approximately equal to the bandwidth while here $U_c\sim5$D is in great contrast to single-band Hubbard model. 
($D$ is the half of the bandwidth in LDA calculation which is about 2~eV near $a\sim2.8\text{\AA}$.)
This is not surprising because the transition is controlled by the energy scale of CT gap instead of $U$.
Second, a $\sim$ 0.5\%D $T_c$ here is also significantly smaller than the $\sim$3\%D in the Hubbard model. 
These quantitative differences all demonstrate that the phase transition is not a one-band type.

\begin{figure}
	\begin{center}
	\includegraphics[width=7cm]{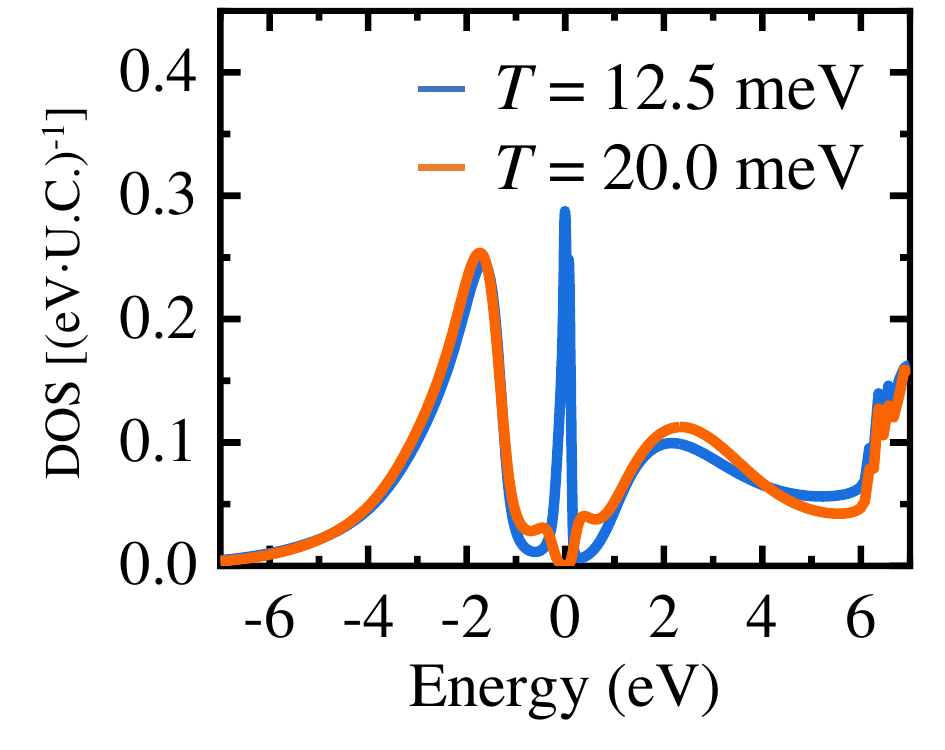}
	\end{center}
	\caption{DOS corresponds to $a= 2.78\text{\AA}$, on the metallic side of the coexistence region. In contrast to the behavior found for a single-band Hubbard model at half filling, here the destruction of metallic quasiparticles leaves behind a pre-formed gap. }
	\label{MIT_fig5}
\end{figure}

In addition to the above difference in the tiny coexistent region, there are more interesting features in the phase diagram in a larger temperature scale.
Figure~\ref{MIT_fig3} shows the value of DOS at chemical potential as a function of lattice constant and temperature.
A significant difference of the above phase diagram compared to the one-band Hubbard model~\cite{Georges1996,Eisenlohr2019} is the existence of an insulator-like phase (in red) even beyond the metallic phase (in light blue)  which should end at $a_{c1}$ in the traditional scenario~\cite{Georges1996,Eisenlohr2019}.
Fig.~\ref{MIT_fig5} shows the DOS at $a=2.78\text{\AA}$ where the system is in metallic phase at low temperature $T=12.5$~meV.
One can find that at high temperature $T=20$~meV, with the destruction of the quasiparticle peak, an insulating gap appears again.

\begin{figure}
	\begin{center}
	\includegraphics[width=\columnwidth]{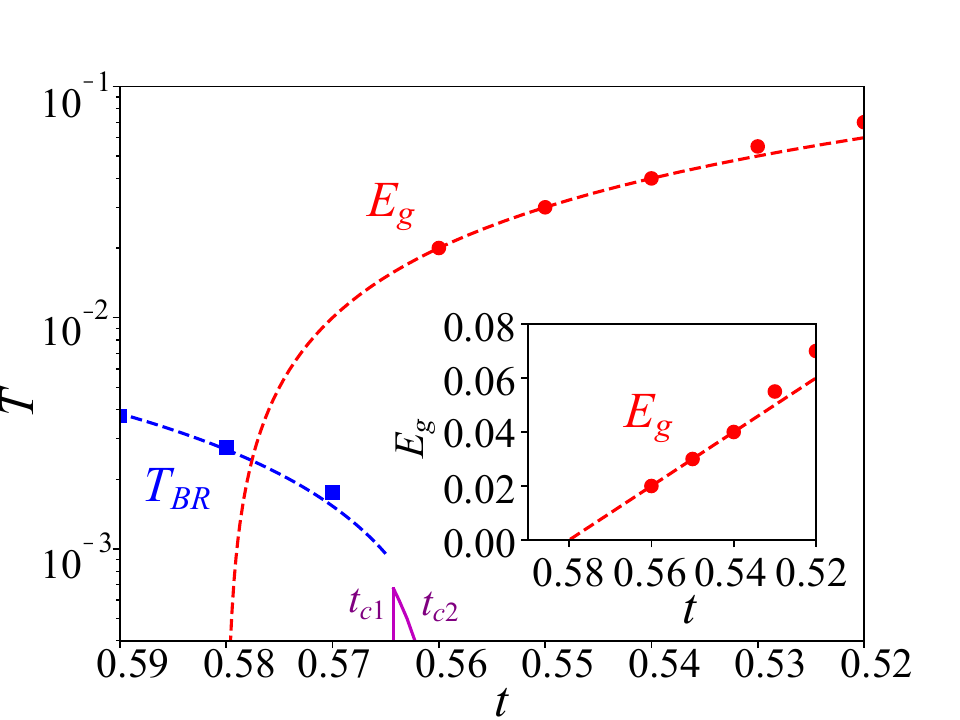}
	\end{center}
	\caption{Phase diagram of a simple two-band model, in the charge-transfer regime (see text). The CT gap $E_g$ (here plotted in the same units as temperature) decreases linearly as the bandwidth increases, but remains sizeable throughout the coexistence region ($t_c \sim 0.563$), extrapolating to zero further on the metallic side (around ($t_g \sim 0.58$). Results shown correspond to $V = 0.217$, and $U =2$. The results are plotted as a function of $t$, which sets the bandwidth. Here we use $\varepsilon_f =1$ as the unit of energy. }
	\label{MIT_fig6}
\end{figure}

Are these features in the phase diagram the specific results of our hydrogen lattice or the intrinsic properties of CT model?
To verify this question, we perform the same calculation on a simplified two-band charge-transfer model without any long-range screening which includes one correlated orbital and one higher-energy orbital, given by the Hamiltonian:

\begin{equation}
    \begin{split}
   H= & \sum_{i\neq j,\sigma}(t-\mu){c}_{i\sigma}^{\dagger}{c}_{j\sigma}+\sum_{i\sigma}(\epsilon_{f}-\mu){f}_{i\sigma}^{\dagger}{f}_{i\sigma} \\
    & +\sum_{i,\sigma}\left(V_{i\sigma}^{*}{c}_{i\sigma}^{\dagger}{f}_{i\sigma}+V_{i\sigma}{f}_{i\sigma}^{\dagger}{c}_{i\sigma}\right)+\sum_{i}+U{n}_{fi\uparrow}{n}_{fi\downarrow},
    \end{split}
    \label{ALM_Hamiltonian}
\end{equation}
where $c^\dagger_{i\sigma}, f^\dagger_{i\sigma}$ create one electron at site $i$ with spin $\sigma$ in two different orbitals, and $n_{fi\sigma}=f^\dagger_{i\sigma}f_{i\sigma}$ is the number operator.
Only $f$ electrons feel the strong intra-atomic repulsion denoted by $U$. $V$ is the coupling between two orbitals, and $\epsilon_f>0$ denotes the charge-transfer energy.
As the phase diagram shown in Figure~\ref{MIT_fig6}, when $U\gg \epsilon_f$, we find a similar strongly suppressed $T_c$ to the hydrogen lattice as well as an insulator phase above the metal phase. 
The similar phase diagram to that of the hydrogen lattice again implies that the phase transition is less relevant to the long-range screening effect.
On the other hand, our results on the simplified CT model indicate these special features in the phase diagram should be the generic behaviors of the charge-transfer system instead of the specific results of this particular hydrogen lattice which needs more investigation in the future.

\bibliography{MainTex}
\end{document}